\documentclass[12pt,draftcls,onecolumn]{IEEEtran}

\usepackage{epsfig}
\usepackage{amsmath}
\usepackage{amsfonts}
\usepackage{amssymb}
\usepackage{subfigure}
\usepackage{booktabs}

\newcommand{\xvv}{{\boldsymbol x}}
\newcommand{\yvv}{{\boldsymbol y}}
\newcommand{\zvv}{{\boldsymbol z}}

\newcommand{\mmse}{{\sf mmse}}

\newcommand{\snr}{{\sf snr}}

\newcommand{\Ac}{{\cal A}}
\newcommand{\Nc}{{\cal N}}
\newcommand{\Xc}{{\cal X}}

\newcommand{\CC}{\mathbb{C}}
\newcommand{\EE}{\mathbb{E}}

\newtheorem{theorem}{Theorem}

\newcommand{\beq}{\begin{equation}}
\newcommand{\eeq}{\end{equation}}

\title{Derivative of BICM Mutual Information}
\author{Albert Guill\'en i F\`abregas and Alfonso Martinez
 \thanks{A. Guill{\'e}n i F{\`a}bregas is with the Department of Engineering, University of Cambridge, Cambridge, CB2 1PZ, UK, e-mail: \tt guillen@ieee.org.}
 \thanks{A. Martinez is with the Department of Electrical Engineering, Technische Universiteit Eindhoven, Postbus 513, 5600 MB Eindhoven, The Netherlands, e-mail: \tt alfonso.martinez@ieee.org.}
}

\begin{document}

\maketitle

\vspace{-10mm}
\begin{abstract}
In this letter we determine the derivative of the mutual information corresponding to bit-interleaved coded modulation systems. The derivative follows as a linear combination of minimum-mean-squared error functions of coded modulation sets. The result finds applications to the analysis of communications systems in the wideband regime and to the design of power allocation over parallel channels.
\end{abstract}

\section{Introduction, Motivation and System Model}
Bit-interleaved coded modulation (BICM) is an attractive solution to construct coded modulation schemes over non-binary signal constellations by concatenating a binary code with a non-binary modulator through an interleaving permutation \cite{Zehavi1992, CaireTariccoBiglieri1998}. We study the sub-optimal non-iterative BICM decoder proposed in \cite{Zehavi1992}. This decoder, when combined with Gray mapping is shown to have a capacity very close to that of coded modulation \cite{CaireTariccoBiglieri1998}.

The derivative of the mutual information as a function of the signal-to-noise ratio (SNR) is becoming increasingly relevant in the analysis and optimisation of communications systems \cite{LozanoTulinoVerdu2006,NguyenGuillenRasmussen2007}. Reference \cite{GuoShamaiVerdu2005} introduces a fundamental relationship between the minimum mean-squared error (MMSE) and the mutual information  in additive Gaussian channels. In particular, \cite{GuoShamaiVerdu2005} obtains that the derivative of the mutual information with respect to SNR is equal to the MMSE in estimating the input given the output of a Gaussian-noise channel. Beyond its own theoretical interest, this relationship has proved instrumental in optimising the power for parallel channels with arbitrary input distributions and in obtaining the minimum bit energy to noise ratio for reliable communication \cite{LozanoTulinoVerdu2006}.

In this letter, we give a formula for the derivative of the mutual information of BICM in Gaussian channels. The formula can be efficiently evaluated numerically without resorting to lengthy Montecarlo simulations, which makes it particularly attractive. In particular, we consider a Gaussian-noise channel
\beq
\yvv = \sqrt{\snr}\xvv + \zvv
\eeq
where $\yvv,\xvv,\zvv\in\CC^L$ are the output, input and noise vectors. We assume that the noise samples $z_\ell$ are independent and identically distributed $\sim\Nc_{\CC}(0,1)$ and therefore $\snr$ denotes the SNR. We assume that $\xvv \in \Xc^L$ are BICM codewords, where $\Xc\subset\CC$ is the signal constellation, with $m=\log_2|\Xc|$. Furthermore, we assume we employ BICM with the decoder of \cite{Zehavi1992, CaireTariccoBiglieri1998}. Then, the input-output mutual information of BICM $I^{\rm bicm}_\Xc(\snr)$ can be expressed as the mutual information of $m$ binary-input continuous-output symmetric parallel channels \cite{CaireTariccoBiglieri1998}
\begin{align}
  I^{\rm bicm}_\Xc(\snr) &=  \sum_{i=1}^{m} I_i(\snr)  =  \sum_{i=1}^{m}\EE \left[\log\frac{  \sum_{x'  \in  \mathcal    X_b^i}e^{-|\sqrt{\snr}(X-x')+Z|^2}}{ \frac{1}{2}\sum_{x'  \in  \mathcal{X}}e^{-|\sqrt{\snr}(X-x')+Z|^2}}\right]
\end{align}
where the sets $\mathcal{X}_b^i$ contain all signal constellation points with bit $b$ in the $i$-th binary labeling position, and the expectation is over the signal constellation points in $\mathcal{X}_b^i$, over the bits $b$ and noise $Z$. All results in this letter assume natural logarithms. The BICM mutual information expression can be very efficiently evaluated numerically using Gauss-Hermite quadratures \cite{abramowitzStegun1964}. The binary-input continuous-output equivalent BICM channel has a noise which is neither additive nor Gaussian, so the results of \cite{GuoShamaiVerdu2005} cannot be directly applied to compute the derivative of the mutual information. 

\section{Main Result}
The main result of this letter is given by the following Theorem.
\begin{theorem}
The derivative of the BICM mutual information is given by
\begin{align}
\frac{d\,I^{\rm bicm}_\Xc(\snr)}{d\,\snr} = \sum_{i=1}^m \frac{1}{2}\sum_{b=0}^1\left(\mmse_\Xc(\snr) - \mmse_{\Xc_b^i}(\snr) \right)
\label{eq:mmse_bicm}
\end{align}
where 
\begin{align}
\mmse_\Ac(\snr) = \EE[|A|^2] - \frac{1}{2^m}\sum_{a\in\Ac}\int \frac{1}{\pi}\left| \frac{  \sum_{a'\in\Ac} a' \,e^{-|\sqrt{\snr}(a-a') + z|^2}}{\sum_{a'\in\Ac} e^{-|\sqrt{\snr}(a-a') + z|^2}}\right|^2 e^{-|z|^2} dz
\label{eq:mmse_awgn}
\end{align}
is the MMSE of an arbitrary input signal constellation $\Ac$.
\end{theorem}
\begin{proof}
Recent results \cite{BrannstromRasmussen2007,MartinezGuillenCaireWillems2007}, introduce the following alternative expression for the BICM mutual information
\beq
I^{\rm bicm}_\Xc(\snr) = \sum_{i=1}^m \frac{1}{2}\sum_{b=0}^1 \left(I^{\rm cm}_\Xc(\snr) - I^{\rm cm}_{\Xc_b^i}(\snr)\right)
\label{eq:bicm_mi_new}
\eeq
where $I^{\rm cm}_\Ac(\snr)$ is the mutual information for coded modulation over signal constellation $\Ac\subset \CC$. Combining \eqref{eq:bicm_mi_new} with the fundamental relationship between mutual information and MMSE \cite{GuoShamaiVerdu2005} 
\beq
\frac{d\,I^{\rm cm}_\Ac(\snr)}{d\,\snr} = \mmse_\Ac(\snr)
\eeq
we obtain the desired result,
where $\mmse_\Ac(\snr)$ denotes the MMSE of an arbitrary input signal constellation $\Ac$.
\end{proof}

Remark that $\Ac$ need not have zero mean nor unit variance. The expression for the MMSE \eqref{eq:mmse_awgn} can again be efficiently evaluated using the Gauss-Hermite quadratures \cite{abramowitzStegun1964}. Note that, since the BICM equivalent channel has a noise which is not additive nor Gaussian, Eq. \eqref{eq:mmse_bicm} is not the MMSE in estimating the input bits given the output.

The expression \eqref{eq:mmse_bicm} finds application in power allocation over parallel channels with BICM. Following \cite{LozanoTulinoVerdu2006}, the result can be used to optimise the power required to maximise the mutual information of parallel channels subject to a given power constraint. The result can also be used to minimise the outage probability over nonergodic block-fading channels \cite{LozanoTulinoVerdu2006,NguyenGuillenRasmussen2007}.

\begin{figure}[htbp]
  \centering 
  \includegraphics[width=0.8\columnwidth]{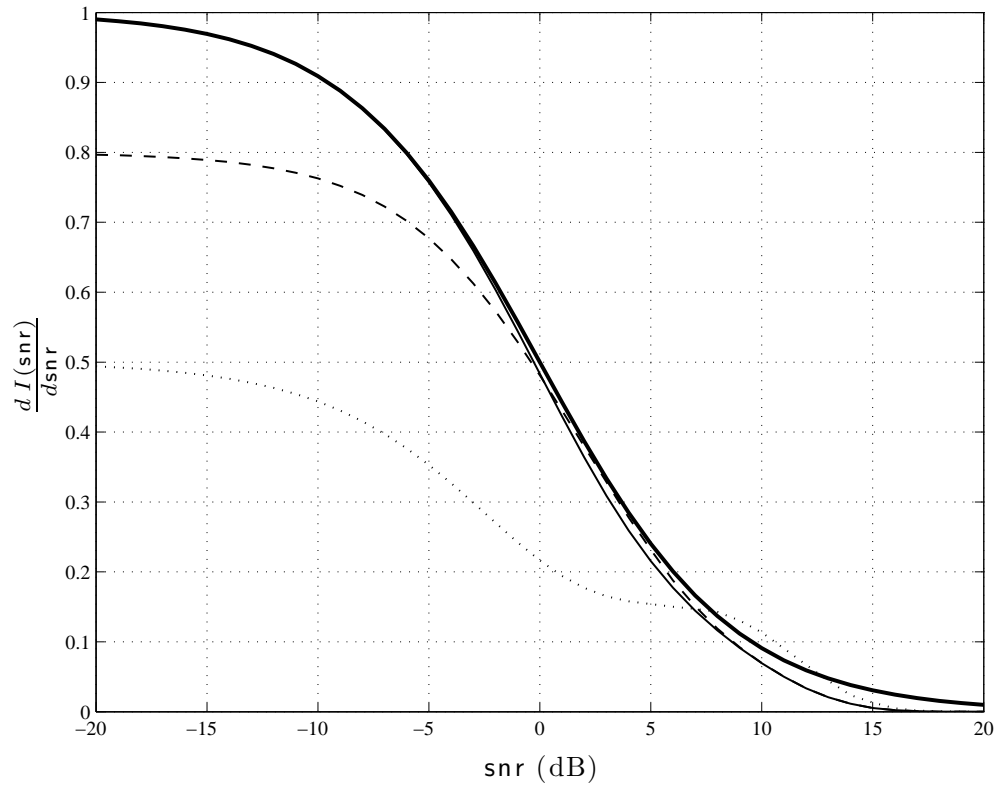}
  \vspace{-2mm}\caption{Derivative of the mutual information for Gaussian inputs (thick solid line), $16$-QAM coded modulation (solid line), $16$-QAM BICM with Gray mapping (dashed line) and $16$-QAM BICM with set partitioning mapping (dotted line).}
  \label{fig:mmse_bicm_16qam}
\end{figure}

Figure \ref{fig:mmse_bicm_16qam} shows an example of the computation of the derivative of the BICM mutual information with $16$-QAM modulation. The MMSE for Gaussian inputs and coded modulation with $16$-QAM are shown for comparison. For Gray mapping (dashed line) we observe a very good match at high $\snr$. At low  $\snr$, the behaviour is given by \cite{MartinezGuillenCaireWillems2007}
\beq
\lim_{\snr\to 0} \frac{d\,I^{\rm bicm}_\Xc(\snr)}{d\snr} = \sum_{i=1}^m \frac{1}{2}\sum_{b=0}^1\left| \EE_{\Xc_b^i}[X]\right|^2.
\eeq
In order to show this result, one may use the methods in  \cite{MartinezGuillenCaireWillems2007} or use the fact that
\beq
\lim_{\snr\to 0} \mmse_\Ac (\snr) = \EE[|A|^2] - |\EE[A]|^2
\eeq
and inserting it into \eqref{eq:mmse_bicm}.

\section{Conclusions}
We have given an expression for the derivative of the mutual information of BICM, which is a linear combination of MMSE functions for coded modulation. The expression can be easily evaluated using numerical integration, and is instrumental in analysing the behavior of BICM in the wideband regime, as well as in deriving optimal power allocation schemes for parallel channels with BICM.

\bibliographystyle{IEEE}

\end{document}